\documentclass[aps,amsmath,prb,a4paper,showpacs]{revtex4} 
\usepackage{graphicx}

\newcommand{\GG}{\breve{G}}

\newcommand{\beq}{\begin{equation}}
\newcommand{\eneq}{\end{equation}}
\newcommand{\bea}{\begin{eqnarray}}
\newcommand{\enea}{\end{eqnarray}}
\newcommand{\bean}{\begin{eqnarray*}}
\newcommand{\eean}{\end{eqnarray*}}

\newcommand{\met}{\frac{1}{2}}

\newcommand{\h}{\hbar}
\newcommand{\bd}{\begin{displaymath}}
\newcommand{\ed}{\end{displaymath}}

\newcommand{\ww}{\omega}
\newcommand{\DD}{\Delta}
\newcommand{\ug}{\underline{\hat{g}}}
\begin{document}
%
\title{Coherent response of a low $T_c$ Josephson junction to an
 ultrafast laser pulse}
\author { P.Lucignano $^{1,2}$}
\author{ G. Rotoli $^{1,3}$}
\author{ E.Santamato $^{1,2}$} 
\author{A. Tagliacozzo $^{1,2}$}
\affiliation{$^1${\sl Coherentia} - Istituto Nazionale di Fisica della Materia
(INFM), Unit\'a di Napoli}
\affiliation{ $^2$ Dipartimento di Scienze Fisiche
Universit\`a di Napoli        "Federico II ", Monte S.Angelo - via
Cintia, 80126 Napoli, Italy}
\affiliation{$^3$ Dipartimento di Energetica,
Universit\`a di L'Aquila, Localit\`a Monteluco, 67040 L'Aquila,
Italy}
\date{\today}
%
\begin{abstract} 
By irradiating with a single ultrafast laser pulse a superconducting
electrode of a  Josephson junction it is possible to drive
the quasiparticles (qp's) distribution strongly out of equilibrium.  The
behavior of the Josephson device can, thus, be modified on a fast time
scale, shorter than the qp's relaxation time. This  could be very
useful, in that it allows fast control of Josephson charge qubits and,
in general, of all Josephson devices. 
If the energy released to the top layer contact $S1$ of the junction 
is of the order of $\sim
\mu J$, the coherence is not degradated, because the perturbation is very fast.
Within the framework of the quasiclassical Keldysh Green's function
theory, we find that the order parameter of $S1$ decreases. 
We study the perturbed dynamics of
the junction,  when the current bias is close to the critical current, 
by integrating numerically its classical equation of
motion.  The optical
ultrafast pulse can produce switchings of the junction from 
the Josephson state to the voltage state. 
The switches can be controlled by tuning the laser
light intensity and  the pulse duration  of the Josephson junction.
\end{abstract}

%
\maketitle
%
\section{Introduction}
The characteristic frequency in the dynamics of a Josephson junction
(JJ) is the so-called Josephson plasma frequency $\omega _{pJ}$ 
(e.g.$10 \div 100 GHz$). Coupling of a JJ to a microwave field
leads to the well known lock-in conditions, which show up as Shapiro
steps in the I/V characteristic. On the other hand, photo-response to
radiation in a superconductor induces heat relaxation (bolometric
effect \cite{barone}) and non equilibrium generation of quasiparticles
(qp) \cite{testardi,parker}.  Both phenomena are extensively studied
since they are relevant for the fabrication of fast and sensitive
detectors.  The models used are phenomenological
\cite{owen,ivlev,ser,lindg,semenov}, mainly involving different
temperatures associated to separate distributions of electrons and
phonons out of equilibrium.\\
\begin{figure}[h]
        \includegraphics[width=\columnwidth ]{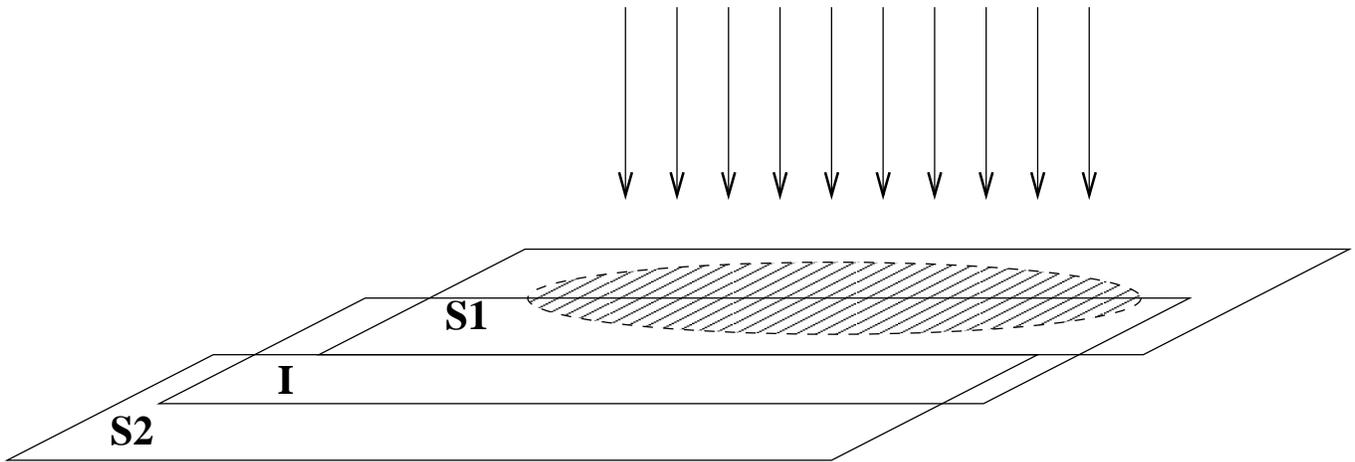} \caption{Sketch
        of the Josephson junction exposed to a laser radiation pulse.}
        \label{fig0}
\end{figure}
Recently, laser light with pulses of femtosecond duration $ \tau _c
\in (10^{-14}s,10^{-13}s )$ has become available, as a source to test
the photo-response of a JJ \cite{lindg}.  Ultrafast pulses can be
extremely useful, in that they allow studying an unexplored regime in
non equilibrium superconductivity.  Indeed, photon absorption, by
creating electron-hole (e-h) pairs at very high energies, drives the
quasi-particle (qp) energy distribution out of equilibrium during the
time $\tau_c$.  The qp non-equilibrium distribution depends on the
energy relaxation time parameter $\tau _E $, defined as the time by
which a `hot' electron is thermalized by repeated scatterings with
other electrons or phonons.  The process involves generation of many
qp's during energy degradation, until the system relaxes back to the
equilibrium distribution function $ n_o (\omega )$. This time scale is
determined by the electron-electron interaction time $\tau_{e-e}$ and
the electron-phonon interaction time $\tau_{e-ph}$, which are strongly
material dependent \cite{kaplan}, ranging from $4\cdot10^{-7} s$ for
$Al$ to $1.5 10^{-10}s$ for $Nb$.  In this work we analyze the
possibility that, keeping temperature quite low, ultrashort laser
radiation induces direct switches out of the Josephson conduction
state at zero voltage, due to coherent reduction of the critical
current $J_c$.

There are many reasons for the switching from the zero to the
resistive state in a Josephson junction. Among these, thermal escape
\cite{thermal}, quantum tunneling \cite{quantum}, latching logic
circuits \cite{latching} and pulsed assisted escape\cite{pulse}. A
clear cut discrimination between different mechanisms can be difficult
to achieve.  In our case quantum escape is ruled out because the
temperature is not expected to be low enough.  Also, we assume that there
is no external circuit to induce switching and re-set of the zero voltage
state as in latching logic elements.

Pulsed assisted escape is a generic term for a large class of
phenomena including in principle bolometric heating of the junction
which is re-set in relatively slow times \cite{testardi}.  Production
of quasiparticles generated by $X$-ray radiation has been studied up
to recently\cite{barone2,ovchinnikov}.  A cascade follows, which
increases the number of excitations and lowers their energy down to
the typical phonon energy $\omega _D$ in a duration time, which is of
the order of the nanoseconds. Subsequently, qp's decay by heating the
sample.  However, the power of the laser can be reduced enough and
both the substrate and the geometry can be chosen such that the energy
released by the radiation on the junction can be small. On the other
hand, appropriate experimental conditions can make the time interval
between two pulses long enough, so that the bolometric response is
negligible.
 
Generally speaking, junctions are more sensitive to pulses especially
when their harmonic content is close to $\omega_{pJ}$, but this is not
our case.  In fact the laser carrier frequency ($\Omega \sim 100 THz$)
is quite high compared to $\omega _{pJ}$ and we consider the case
$\Omega >> \tau _c^{-1} > \omega_{pJ} > \tau _E^{-1}$, what implies that little
relaxation takes place during the duration of the pulse.  $\tau _c $
should also be shorter than the pair-breaking time $\h/\Delta_o \sim 1
\div 5 ps$. Here $\Delta_o $ is the unperturbed gap parameter.
Our approach assumes that, on a time scale intermediate between the
pulse duration and the relaxation time $\sim \tau _{E}$, the order
parameter of the irradiated superconductor is sensitive to the
non-equilibrium qp distribution, which modulates it coherently till it
switches out of the zero voltage state.

To analyze the dynamics of the order parameter and the way how the
latter affects the Josephson current, we adopt a non-equilibrium
formalism based on quasiclassical Green's functions \cite{rammer,belzig}.
The quasiclassical approach has been mostly used in the past in
connection with the proximity effect \cite{belzig}, as well as with
non-equilibrium due to other space inhomogeneity
conditions\cite{amin}.  As far as we know, this is the first time that
its extension to non-equilibrium in time is applied to a coherent
response after an ultrafast laser irradiation.

The quasiclassical approximation to the Gorkov equations, is obtained
by averaging over the period of the optical frequency $\Omega $, which
is a fast time scale\cite{noi}.  Our equations include the physics of the
cascade process, which occurs when one focuses on the kinetics of the
qp diffusion. A  kinetic equation   approach  to  the steady  state 
non equilibrium  qp  distribution,  including phonon  scattering has 
 been  developed in  ref.\cite{scalapino}, for  light  irradiation, 
 mostly  in  the microwave  range.   The  cascade  regime is extensively 
discussed in Ref.\cite{ovchinnikov}, however it will not be specifically 
addressed  here.

Instead, if the switching  of the irradiated superconductor due to the
ultrafast pulse takes place prior to the occurrence of qp relaxation,
an approximate solution of the dynamical equations can be derived,
which describes an istantaneous response of the order parameter.

We take a low $T_c$ JJ with an $s$-wave order parameter as the reference
case (e.g. a high quality $Nb$ or $Al$ junction) and $T<<
T_c$. The optical penetration depth of the laser light $
\lambda_\delta$ in the topmost superconductor exposed to radiation
$S1$ is assumed to be shorter than its thickness, so that any
modification induced by the radiation field only involves $S1$ itself
\cite{testardi} (see. Fig.(\ref{fig0})).  In a small size JJ the
spatial variation of the order parameter along the lateral dimension
of $S1$ is not taken into account, except when the qp diffusion
process cannot be ignored.

We consider just one pulse of given duration $\tau _c$ which releases
the energy ${\cal{E}}$ per pulse, by exciting $e-h$ pairs and by
creating a non equilibrium distribution of qp's. A related
dimensionless quantity $q$, as defined in Eq. (\ref{qq}), parametrizes
the strength of the perturbation due to the radiation. The
perturbation is assumed to be small so that only the lowest order in
the expansion in $q$ is retained.  This allows us to derive a
temporary reduction of the order parameter $\Delta $ induced by the
pulse, as shown in Fig.(\ref{fig1}).  We do not give a detailed
description for the relaxation of the non-thermal qp distribution in
the irradiated superconductor.  The self-energy terms corresponding to
this process require further analysis. According to the Eliashberg
formulation \cite{kaplan} these terms affect the quasiparticle
amplitude $Z(\omega)$ introducing changes in the phonon distribution
and retardation in the response. Nevertheless, we expect that these
self-energy terms become effective only on a longer time scale after
the laser pulse.  Our equations pinpoint a non retarded evolution of the
order parameter prior to relaxation, which implies a reduction of the
critical current. This shows that the coherent modulation of the gap
parameter can produce switching of the junction out of the Josephson
state.

The switches are studied numerically by solving the classical equation
fo motion of a current biased JJ with current $J$ close to the
critical curent $J_c$, during the excitation process.  After the
switching the dissipation in not treated selfconsistently: a standard
dissipation, typical of thermal equilibrium, is assumed in the JJ
dynamics, by adding a conductance term in the numerical simulation.
We stress that the assumed model for dissipation determines the actual
qp branch of the I-V characteristics, but does not affect
substantially the switching probability. The switching from the
Josephson state to the voltage state in the parameter space
$(q,\tau_c,J/J_c)$ is reported in Fig.\ref{fig5}. Interestingly, we
find that for fixed value of $q$ and $J/J_c$ there is an optimum pulse
duration to achieve the strongest sensitivity of the junction to the
switching process.  We show that this is due to the way how the non
equilibrium qp's distribution affects the pairing in $S1$.  The paper
is organized as follows.

In Section 2 we calculate the non equilibrium qp distributions
immediately after the pulse, prior to relaxation.

In Section 3 we introduce the time dependent quasiclassical Keldysh
Green functions formalism extended to the time domain, using the
general frame given in appendix A. We calculate the  correction to the
single particle
propagators to first order in $q$. A correction to the gap
and to the Josephson critical current follows due to the laser pulse.

In Section 4 the dynamics of the JJ after the pulse is simulated
numerically. Finally,  a summary of our results is given in Section 5.

Appendix A collects the formulae of the quasiclassical approximation
in the non equilibrium Keldysh theory, which are used in the core of
the paper.  In Appendix B we derive the kinetic equation for the non
equilibrium distribution function which drives the relaxation of the
system. The equations of motion for the quasiclassical retarded Green
function is reported in appendix C, while the  equation of motion for the
advanced and Keldysh Green function can be derived in the same way .

\section{The non-equilibrium qp distribution}

\subsection{Non-equilibrium electron-hole pair  excitations induced by
optical irradiation}

The optical frequencies ($\Omega \sim 100 THz$) building the
wavepacket of the laser pulse excite $e-h$ pairs at high energies.  As
explained in the introduction the non-equilibrium arising from the
alteration of the qp distribution has a relaxation time $\tau _E$
which is long compared to the optical period: $\Omega \tau _E >>1 $.
In addition to this, the duration of the pulse $\tau_c\sim \omega _c
^{-1} $ is even shorter than the pair breaking time, so that we expect
that, in our case, dissipative phenomena do not affect the coherence of the
superconductor on the  time scale $\tau_c$.

Qp's are generated as if the metal were normal, because
superconductivity doesn't play any role in their excitation at large
energies.  They propagate according to something very much like the
free particle time-ordered Green's function $g_o $ (from now on we put
$\hbar =1 $):
\beq
g_o^T(k ;t-t' ) \equiv 
-i  \langle T[c_k(t) c_k^\dagger (t') ] \rangle = -i e^{-i\xi (t-t') } 
\left [ ( 1- n_k ) \theta (t- t' ) - n_k  \theta (- t+ t' )  \right ]\:.
\label{green} 
\eneq
Here $\xi $ is the qp energy with momentum $k$ and is measured from
the chemical potential $\mu$. $n_k$ is the qp distribution function.
We assume that $e-h$ symmetry is conserved in the excitation process, so
that $\mu$ is not altered with respect to its equilibrium value.
    
The equation of motion for the Green's function $\tilde g$ in the
presence of the radiation field is:
\beq
\left (i  \frac{\partial}{\partial t} - \frac{1}{2m}[\vec \nabla _r - 
\frac{e}{c}
\vec A(\vec r,\vec R , t)]^2 + \mu \right )\: 
\tilde g (\vec r, \vec R ,t,t')= 
 \delta(r) \delta(t-t') 
\label{bg1}\; ,
\eneq
$\vec r$ is the relative space coordinate,  while $\vec R$ is the center of
 mass  coordinate. 
The vector potential is a wave-packet centered at frequency $\Omega$
according to: 
\beq
\vec{A}(\vec r, \vec{R},t)=\sum_{\pm}\sum _{\vec p} \vec{a}_\pm
(\vec p, \vec{R},t)e^{\mp  i ( \vec {p}\cdot \vec{r}  -   \Omega t)} \; .
\label{potvet}
\eneq
Here $\vec{a}_{\pm}$ are slowly varying 'envelope' functions of $\vec R$ 
on the
size of the irradiated spot and on the  time scale $\Omega^{-1}$.  
We look for solutions of eq.(\ref{bg1}) in the form:
\beq
\tilde g (\vec{r}, \vec{R};t,t') = g(\vec{r}, \vec{R};t,t')
 + \sum_{\pm}\sum _{\vec{p}}g ^{\pm}(\vec{p}, \vec{R},t,t')
 e^{\mp i ( \vec {p}\cdot \vec{r}  -   \Omega t)}\;\; ,
\label{ciccia}
\eneq
where $g $ and $g^{\pm}$ are slowly varying functions of $
\vec{R}$ and $t$ on the same scales. A similar expansion can be
done w.r.to the variable $t'$.  Following Eq. (\ref{ciccia}), a
decomposition of eq.(\ref{bg1}) into harmonics arises\cite{noi}.  We define the
zero order harmonic equation as the one that does not contain
exponentials $e^{\pm i \Omega t}$.  By averaging over a period $\Omega
^{-1} $ we neglect harmonics of order two, or higher. This amounts to
include one photon excitation processes only, with released energy
${\cal{E}}$.  Some  extra  details  can  be found  in  Appendix A:
\beq
\left (i \frac{\partial}{\partial t} + \frac{1}{2m} \nabla ^2_r +
 \frac{e^2}{mc^2} \sum _{\vec {p}',\vec{p}''} \vec{a}_+
(\vec {p}', \vec{R},t) \vec{a}_-
(\vec {p}'', \vec{R},t)e^{  i ( \vec {p}'-\vec{p}'') \cdot \vec{r} }\: 
+\mu  \right ) 
 g (\vec r, \vec R ,t,t')= 
 \delta(r) \delta(t-t') 
\nonumber\; .
\eneq
 
Fourier   transforming w.r.to $\vec r$ ( $\vec r \to \vec p \to  \xi  $) 
 we have:
\beq
\left (i \frac{\partial}{\partial t} - \left (\frac{p^2}{2m}
 -\mu \right )  \right )  g (\vec {p}, \vec R ,t,t') +
 \frac{e^2}{mc^2} \sum _{\vec {p}',\vec{p}''} \vec{a}_+
(\vec {p}', \vec{R},t) \vec{a}_-
(\vec {p}'', \vec{R},t)\:  
 g (\vec {p}'-\vec {p}''-\vec {p}, \vec R ,t,t')=  \delta(t-t') 
\label{bg21}\;.
\eneq
The  radiation  field  generates and annihilates  high   energy 
$e - h$  pairs.  Hence  we assume that the  forcing  term  conserves the
total  impulse, $  \vec {p}'+\vec {p}'' \sim  0 $, but   
$\vec {p}'-\vec {p}''$ transfers an  energy $2\xi  $ to  the electrons. 
Therefore we take   the coupling term  in  the  Hamiltonian as:
\beq
\frac{e^2}{m c^2  }  \vec{a}_+
(\vec {p}', \vec{R},t) \vec{a}_-
(\vec {p}'', \vec{R},t )   \to 
 q \:
\frac{\ww _c }{\sqrt{\pi}} e^{-\met \omega _c^2 t^2} \:  \delta 
(2 \xi  + \xi ' ) \:\delta ( \vec {p}'+\vec {p}'' ) \: .  
\label{defq}
\eneq
A  Gaussian shaped time dependence  has been chosen for the 
pulse with  half-width $\omega _c^{-1}$, while the  space dependence
has been   neglected  for  simplicity. In Eq. (\ref{defq}) the dimensionless 
quantity  appears:
\beq 
q\sim\frac{e^2}{c }\: \frac
{\omega_D }{\Omega } \: \;  \frac  {{\cal{E}}} {m \Omega^2 R^2_o }\:,
\label{qq}
\eneq
Where $R_o$ is the laser spot (see Eq.\ref{brown}).
Here  the number of excited  $e-h$ pairs  is
$\sim\Omega/\omega_D$, with $\ww_D$ the Debye energy.
Experiments \cite{testardi,lindg} show that the energy released by the
pulse can be very low, so that we will always expand in $q$.  In fact,
while in the case of an $ rf $ radiation $q \sim 1 $, in the case of a
femtosecond laser pulse  $q\sim 0.01\div 0.1 $, corresponding to
a fraction of $\mu J $ released per pulse on the superconducting
surface of $ \sim 100 \mu m ^2$.

The zero
order harmonic equation,  Eq. (\ref{bg21}),  becomes :
\beq
\left (i \partial _t  -\xi  \right )   g (\xi; t,t' )  +
  q \frac{\omega _c}{\sqrt{\pi}}
 e^{-\met \ww _c^2 t^2 }   g (- \xi; t,t' ) = \delta (t-t') 
\label{bg22})\;.
\eneq

To derive the non-equilibrium correction to the qp distribution
function, the kinetic equation  \ref{ovA2} should be solved.
In place of this we proceed  in  this  work  in an heuristic way.
Our  approach  lacks mathematical rigour,  but   singles 
out  directly  the role  of the laser induced $e-h$ excitations  
at  frequencies $( \Omega -\omega _c , \Omega +  \omega _c )$.
Our result is valid
in the limit of large $\xi $'s and zero temperature, before
relaxation takes place.

We solve Eq(\ref{bg22})  for the retarded Green's function for $t>0$  and 
$-t' \sim 0^+$ by   truncating the Dyson  equation to lowest order  in $q$:
\bea
 g^R (\xi; t,t') = g_o^R (\xi; t-t') - q \frac{\omega _c}{\sqrt{\pi}}
 \int dt'' g_o^R(\xi; t,t'') e^{-\met \ww _c^2 {t''}^2 } g_o^R (- \xi;
 t''-t' ) \nonumber\\ = g_o^R (\xi; t-t') +i q \frac{\omega
 _c}{\sqrt{\pi}}
\int  _{-\infty}^{+\infty}  \frac{d\omega}{2\pi}
    \frac{e^{-i \omega t}}{\omega- \xi  +i0^+} \int _{0^+}^{+\infty}
 dt \:   e^{i (\ww -  \xi ) t } 
 e^{-\met \ww _c^2 t^2 }\label{sopra}\;,
\enea  
where $g_o^R (\ww ) = \{ \ww -\xi + i 0^+ \}^{-1} $ is the Fourier
transform of the retarded Green's function.  The time integral can be
expressed in terms of the function $ w[z] \equiv e^{-z^2} \: erfc\:
(-iz) $. If we now approximate Eq. (\ref{sopra}) by evaluating $w[z]$
only at the pole and use the the integral representation of the step
function:
\beq
\theta (t)= \frac{e^{i z t}}{2 \pi i} \int _{-\infty}^{+\infty} d\ww
\frac{e^{- i \omega t}}{\omega-z} \;\;\; for \;\; \Im m z < 0\:\:\: ,
\eneq
 the correction $ \delta g^R ( \xi , t)$ to $g^R$ for $\xi >> 0 $,
which includes the non  equilibrium  qp 's distribution, 
is: \beq \delta g^R ( \xi , t, 0^-)= - \frac{q}{\sqrt {2}} e^{-i (\xi
-i0^+) t}\: \: w\left [- 2\xi /(\sqrt {2}\ww _c ) \right ] \:\: \theta
(t ) ,
\label{corrg}
\eneq
 From eq.(\ref{corrg}) we obtain the time
ordered Green's function for $t>0 ,\: t' = 0^- $ and $\xi >0$:
\beq 
g ( \xi , t>0 , 0^- )= (-i ) e^{- i (\xi -i0^+) t}\: \theta (t )\:
\left ( 1 -  \frac{q}{\sqrt{2}}   \:
  \: \rho\left [  2\xi   /(\sqrt {2}\ww _c ) \right ] \right )   ,
\label{gcorr}
\eneq
with
\beq
 \rho [ x ] \equiv e^{- x^2 } \frac{2 }{\sqrt \pi } \int _0^ { x } ds
 \: e^{ s^2 } \:\: .
\eneq
$ \rho [ x ] $ increases linearly with $x$ and it decreases slowly, as
$1/x$, at large arguments.  In Eq. (\ref{gcorr}) we have neglected $\Re
e [w] $ because $|\xi | >> 0 $. 

Eq. (\ref{gcorr}) is to be compared with the free propagating time
ordered Green's function of eq.(\ref{green}) for the same time
arguments.  Comparison yields the amount by which the distribution
function is driven out of the equilibrium:
\beq
\delta n (\xi )  \approx  \frac{q}{  \sqrt {2}} 
 \: \rho \left [ 2\xi /(\sqrt {2}\ww _c ) \right ] \;\;\; for \;\; |
 \xi | >> \DD _o \:\:\: .
\label{dnx}
\eneq
Note that the expression of Eq. (\ref{dnx}) changes sign according to $
sign\; \{\xi \}$. This stems from the assumed $e-h$ symmetry.  In turn
this implies that no charge imbalance occurs.

Eq. (\ref{dnx}) can be considered as the non equilibrium distribution   for
qp 's starting at the time of the pulse $t\sim 0^+ $.
  
In the absence of relaxation, a change in the available qp density of
states follows.  Because $(g^R(\xi ,\ww ) )^* = g^A(\xi ,\ww ) $ if
$\ww$ is real, the correction to the density of state $\delta \nu
(\xi)$ is:
\beq
\delta \nu (\xi)  = -\frac{1}{\pi} \left ( 
    \delta g^R - \delta g^A \right ) ( \xi , 0^+) \approx
\frac{q}{\pi  \sqrt {2}}
  \: \rho\left [ 2\xi   /(\sqrt {2}\ww _c ) \right ] 
 \;\;\; for \;\; \xi  > 0\:\:\: . 
\label{dens}
\eneq
The first stages of the relaxation process involve the inelastic
diffusion of qp's in the medium which is qualitatively discussed in
the next section.

\subsection{Inelastic diffusion   of the qp's at initial  times }
Let us discuss shortly what was neglected in the derivation of the
change in the equilibrium qp distribution $ \delta n(\xi ) $ given by
Eq.(\ref{dnx}).
 
The single particle Green's function $g (p,R,t,t' ) $ is assumed to be
a slowly varying function of $ (t+t' )/2 = \overline t $ and a fast
varying function of $t -t'$. Fourier transforming w.r.to the latter
variable ( see Appendix A ) there is an $\omega $ dependence even in
the stationary case ( i.e.  with no $\overline t$ dependence). This
$\omega $ dependence is determined by the frequency dependent
Eliashberg $e-ph $ coupling $\alpha ^2 F(\omega ) $ and is contained
in the $e-ph $ self-energy $M_{e-ph} (\omega ) $ \cite{grimvall}.
Accordingly, the complex qp renormalization parameter $ Z_n (\omega )
$ is defined by $[1- Z_n (\omega ) ] \omega = M_{e-ph} (\omega ) $.
In our derivation we have not included the self-energy, so that we are
implicitly taking $ Z_n (\omega ) \to 1 $, what applies for large
$\omega ( \sim \Omega ) $, prior to relaxation.

Moreover, because the system is in the superconducting state, we
should have dealt with the corresponding superconducting parameter $
Z_s (\omega )$.  The latter is derived
 together with the complex gap parameter $\Delta ( \omega ,
\overline t ) $ with $\Delta ( \Delta_o , \overline t=0^- )=\Delta_o $  
from the coupled Eliashberg equations ( we drop the overline on $t$ in
the following ).

The procedure of averaging over the fast time scale $\Omega ^{-1}$
singles out two frequency components of $ \Delta ( \omega , t ) $ and
$ Z_s (
\omega , t ) $: $\omega = \Delta _o$ and $\omega=\Omega  $ as a consequence retardation arises from frequencies up to $ 10\omega_D$ is neglected.  
$\Omega$ is so large that $ Z_s $ and $ Z_n $ do not differ sizeably.
In fact, their real parts differ by a quantity of the order of $
(\omega _D \Delta _o /\Omega ^2 )^2 \ln (\omega _D / \Delta _o ) $.  $
\Delta ( \Omega , t ) $ itself is expected to be so small that it can
be neglected alltogether.  Indeed, in connection with eq. \ref{kcomp}
of Appendix A we do not discuss the self-energy terms.  Of course this
approximation breaks down on the time scale of $e-ph $ relaxation.
 
Let us now discuss the $t$ dependence.  The
 equation of motion  for the qp distribution function $n(\vec{R},t)$ is
derived in Appendix B, where we take $n_T =0$, because we neglect
charge imbalance corrections.
 
In averaging over a few optical periods the kinetic equation for
$\delta n _L $, the electric field $\vec E$ averages to zero. The qp
relaxation is governed by the collision integral $I[n(\vec{R},t);t]$
which describes the inelastic processes. In Ref.\cite{ovchinnikov} the
cascade of the $e-h$ excitations due to inelastic scattering is studied
in detail.  Two stages occur. In the first stage e-e interactions
multiply the number of excited qp's in the energy range from $\Omega$
to $\omega _D $ which is taken as the cutoff energy of the pairing
interaction. This happens in a time interval short w.r.to the pulse
duration ($\sim 10 ^{-14} s $). In the second stage a much slower
relaxation process takes place, by which the energy of the qp's
reaches $\DD $.  This process involves electron-phonon scattering on a
time scale $ \hbar \omega_D^2/\DD ^3
\sim 10^2 ns$ which is much larger than any time scale in our problem.
Here we will leave this stage aside.  In the time interval we are
concerned with, we have little relaxation and the energies involved
are $\omega >> \DD $.

According to Eq. (\ref{funh}) the distribution function prior to
relaxation deviates from the equilibrium value by the quantity $\delta
n_L = -2 \delta n (\xi ) $ given by Eq. (\ref{dnx}). There is no
explicit dependence on $\omega$ in our correction, becauise
retardation is nelgected. Still qp's diffuse in space inside the
junction over a characteristic distance $R_o \sim \sqrt{ D
\tau _{e-e} } $, where $D$ is the diffusion coefficient. Hence
\beq
\delta n_{L}(\xi, \vec{R},t)=
- 2  \frac{R_o^2}{R_o^2+ D t} \:
\delta n (\xi )\:
e^{\left \{-\frac{R^2}{R_o^2+\frac{D} t }\right\}} \:\:\:  .
\label{brown}
\eneq 
For relatively large times $ \tau_{e-ph}>>t >> R_o^2/D $ we will
ignore the spatial dependence by putting $R = 0$.  This is the first
step of a perturbative analysis of the non equilibrium distribution
functions.

\section{Changes   of the superconductive 
properties on the time scale $\ww _c ^{-1} $}

\subsection{The correction  to the gap  parameter}

In this Subsection we derive the Keldysh Green's function in the
presence both of a time dependent gap $\DD (t)$ and of a
non-equilibrium qp distribution as given by Eq. (\ref{brown}).  We
assume weak coupling superconductivity and we neglect here the
frequency dependence of the $e-ph$ coupling parameter $\lambda $. This
follows from the neglecting the retardation effects mentioned in
Sec. IIb on time scales much faster than the $e-ph $ relaxation
time. From the Keldysh Green's function $\hat g ^K$ (where the hat
denotes matrix representation in the Nambu space, see appendix A) we
recalculate the gap self-consistently, according to the formula:
\beq
\DD(t)=- \frac{\nu(0) \lambda}{4} \int_{-\infty}^{\infty}
<f^K({\vec p/|\vec p|},\omega,t)>_{\vec v_F} \;
d \omega \;.
\label{gap}
\eneq  
The average over the direction of the momenta on the Fermi surface is
indicated.  The Keldysh Green's function in thermal equilibrium is:
\beq
\hat  g_o^K  =\tanh \frac{\beta \ww }{2} \left(\hat g^R - \hat g^A\right)\:.
\label{gok} 
\eneq
Out of equilibrium we use the definition:
\beq
 \hat g^K =   \hat g^R \hat h - \hat h \hat g^A    \:\:  .
\label{ggk}
\eneq
However, $ \hat h $ defined in appendix B is here diagonal, because we
assume that no charge imbalance arises. Hence, up to first order in $q$,
\beq
\delta \hat g ^K \approx n^0_L(\hat g^R_{ad}-\hat g^A_{ad})
+\delta n_L (\hat g_o^R- \hat g_o^A)\:.
\label{lingk2}
\eneq
 Here $ n_L^o = \tanh (\beta \ww
/2) $ is the equilibrium distribution and $\hat g^R_{ad}-\hat
g^A_{ad}$  is introduced in Appendix C (see Eq.\ref{fg}) and is
discussed in the following.

We now first derive the contribution coming from the second term of
Eq. (\ref{lingk2}).  We start from the outset using Eq. (\ref{brown})
and performing the quasiclassical approximation.  The latter involves
an energy integration:
\beq
\delta n_L \cdot \left ( \hat {g}_o^R - \hat {g}_o^A \right )
\equiv
\frac{i}{\pi}\int _{-\infty}^{+\infty}
d\;\xi \: \delta n_L (\xi ,t) \cdot \left ( \hat {g}_o^R (\xi, \omega
,t) - \hat {g}_o^A (\xi, \omega ,t)\right )\:.
\label{qcl}
\eneq

Using the equilibrium BCS functional forms, the Green's functions
appearing on the diagonal of $\hat{g}^{A/R}$ are \cite{gork}:
\bea
g^R(\xi , \omega)=\frac{u^2_{\xi}}{\omega- E_{\xi} +i 0^+} +
        \frac{v^2_{\xi}}{\omega+ E_{\xi} +i 0^+}\:,\\
g^A(\xi  , \omega)=\frac{u^2_{\xi}}{\omega- E_{\xi} -i 0^+} +
        \frac{v^2_{\xi}}{\omega+ E_{\xi} -i 0^+}\:.
\enea 
The equilibrium values for $u_{\xi}$ and $v_{\xi}$ are:
\beq u_{\xi}^0
= \left (\met( 1 + \frac{\xi }{E}) \right )^\met \: ,
\hspace*{1cm} v_{\xi}^0 =  \left (\met( 1 - \frac{\xi }{E}) \right
)^\met \: ,
\eneq 
with $E = \sqrt{{\xi}^2 + |\Delta_o|^2 }$. From now on we will drop
the subscript in the equilibrium gap parameter
(i.e. $\Delta\equiv\Delta_o$ if no time dependence is indicated
explicitely).  Because the factor $\delta n_L(\xi,t) $ appearing in
eq.(\ref{qcl}), as given by Eq. (\ref{brown}) is odd w.r. to $\xi$
only the second term in $u^2$ and $v^2$ survives, when the integral in
Eq. (\ref{qcl}) is performed.  Let us consider the case $\ww > \DD$
only and specialize Eq. (\ref{qcl}) to its diagonal part.  According
to Eq. (\ref{brown}) we have:
\bea
\Re e\left \{\delta  n_L \cdot g^R_o  \right \}  =
\Re e \left \{  -   \frac{\sqrt {2} i}{\pi}\:   
q(t)  \: \int_{-\infty} ^{+\infty }d\:\xi  \frac{ \xi }{2 E}
  \: \rho\left [  2\xi    /(\sqrt {2}\ww _c ) \right ] 
  \left (\frac{1}{\omega- E+i 0^+} -
            \frac{1}{\omega+ E+i 0^+}\right ) \right \}=
\nonumber \\
=   - q(t)   \: 
  2\sqrt {2} \:  
 \rho\left [ 2(\ww ^2 -|\DD |^2 )^\met  /(\sqrt {2}\ww _c )
 \right ]\:.
\label{qqqq}
\enea
Here we have defined the function $q(t)$:
\beq
q(t) = q\:  \:
 \frac{\pi  R_o^2}{R_o^2+ Dt} \:.
\eneq
Doing similarly  for $g^A$  and subtracting, the  imaginary  part  cancels:
\beq
\delta n_L \cdot \left (  {g}_o^R - {g}_o^A \right )  =
- q(t) 4 \sqrt {2} \: \rho\left [ 2(\ww ^2 -|\DD |^2 )^\met /(\sqrt
{2}\ww _c ) \right ] \; ,\;\; for \;\; \ww > \DD \:\:\: .
\eneq 
 Here  the  largest  contribution  of  the   non-equilibrium  
excitations  arises   from   $ \omega  \sim  \ww _c$. On  the  other  hand 
$ \ww _c$ can  be  larger  or  smaller than  $\omega _D $.

Now we evaluate the correction due to $\hat g^R_{ad}-\hat g^A_{ad}$.
In appendix C we show that an adiabatic solution of the motion
equation of $g^{R,A}$ is possible, in the sense that the functional
dependence on $\ww$ is the same as the equilibrium one, but the gap
parameter changes slowly with time (see Eq. (\ref{fg})):
\beq 
\hat g_{ad}^{R(A)}=+(-)\frac{ \hat
M}{\sqrt{(\omega \pm i 0^+)^2-|\Delta (t) |^2}} \:
\label{fg1}
\eneq
with
\beq
\hat M=\left( \begin{array}{cc}
\omega & \Delta(t)\\
-\Delta(t) ^* &-\omega 
\end{array}\right). 
\eneq
This functional form for the $R/A$ functions is obtained if the $e-h$
symmetry is maintained and if one neglects the diffusion in space-time
which will be mainly important at intermediate times
\cite{ovchinnikov}.\\ This adiabatic approximation in the advanced and
retarded Green's functions allows us to write the Keldysh propagator in
the form:
\beq
g^K=   g_{ad}^{K}- q(t)   4\sqrt {2}
\:   \rho\left [ 2(\ww ^2 -|\DD |^2 )^\met  /(\sqrt {2}\ww _c )
\right ] \:.\label{gggg}
\eneq
To calculate $f^K $ we resort to the analogous of eq.(\ref{ff}) which
is valid for $\hat g^K $: $ f^K = g^K \DD /\ww $.  Hence we have:
\beq
f ^K = f_{ad}^K - q(t) 4 \sqrt {2} \: \frac{\DD}{\ww } \: \rho\left [
2(\ww ^2 -|\DD |^2 )^\met /(\sqrt {2}\ww _c ) \right ] \;\;\; for \;\;
\ww >> \DD \:\:\: . \label{anomalo}
\eneq 
Now we insert Eq. (\ref{anomalo}) into Eq. (\ref{gap}) and consider the
linear correction to the gap of the irradiated contact according to :
\[\Delta(t)= \Delta+\delta\Delta(t)\:.\] 
Here $\delta\Delta(t)$ is the correction  to   the 
unperturbed gap parameter $\DD$ due   to
the radiation. In the limit of zero temperature, up  to 
first order in $q (t)$ and $\Delta/\omega_D$, $\delta \Delta(t)$  is  given by:
\beq
 \frac{\delta \DD (t)}{\Delta} =
 -\frac{q(t)
 4 \sqrt {2}}{ln(2\omega_D/\Delta)-2+{\cal{O}}(\Delta^2/\omega_D^2)} \int_{\Delta }^{\omega_D} \; \frac{  d
 \omega }{\ww} \: \rho\left [ 2(\ww ^2 -|\DD |^2 )^\met /(\sqrt {2}\ww
 _c ) \right ] \: .
\label{delgap}
\eneq 
\begin{figure}[ht]
        \includegraphics[width=\columnwidth]{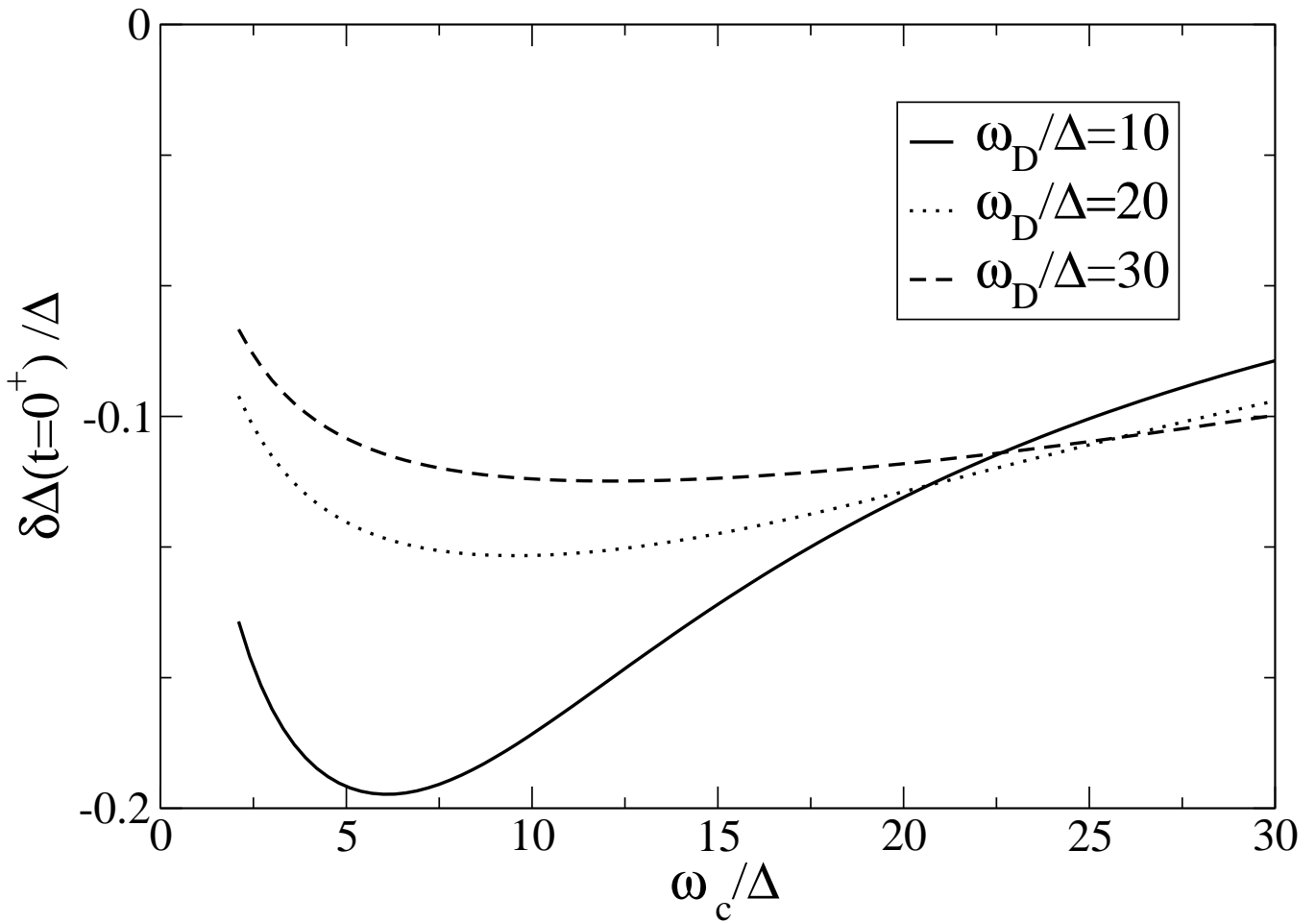}
        \caption{Variation of the gap immediately after the pulse vs
        the inverse of duration of the pulse $\tau_c^{-1}=\omega_c$ in
        units of $\Delta$.  Our approximations break down for very low
        ratios $\omega_c/\DD$ (long pulses). On the other hand, large
        $\omega_c/\DD$ represent very short pulses: this situation is
        unrealistic with the available optical devices.} \label{fig1}
\end{figure}
 It is interesting to
note that the correction arising from the adiabatic dynamics has the
role of renormalizing the coupling $q(t)$ via the prefactor
$-[ln(2\omega_D/\Delta)-2]^{-1}$. This prefactor is negative because
$ln(2\omega_D/\Delta)>2$. Therefore Eq. (\ref{delgap}) shows that the
gap of the irradiated superconductor is decreased due to the non
equilibrium distribution of the qp excitations. 

Eq. (\ref{delgap}) has the same structure as Eq. (14) of ref.\cite{scalapino}.
There is a striking difference however. 
The inverse square root singularity at the gap  threshold,
which   shows  up  in  Eq.(14)  of  ref.\cite{scalapino}
 does  not  appear in Eq. (\ref{delgap}).

 The  inverse    square root 
singularity    originates from  the unperturbed density of
states   at  the  excitation  threshold  and  is  usually  present
in non-stationary  superconductivity \cite{eliashberg}. It  is  responsible 
for  retardation   and  oscillating  tails.
In  our  case, the  gap  threshold 
 plays a little role,  because we do  not  
have  extensive pair-breaking  and q.p. generation 
at energies $\sim  \Delta $.  
Hence, just a  tail  $ \sim 1/\omega $
survives in   the integrand. 

In Fig.(\ref{fig1}) the variation of the gap immediately after the
pulse ($t\sim 0$), is plotted vs the inverse of the pulse duration 
 $\omega_c$ in units of $\Delta$, for different values of $\omega_D$.  
Our approximations are not valid when the
pulse becomes too long (very low values of $\omega_c/\DD$). 
For longer pulses the integrand in Eq. (\ref{delgap}) has a narrow peak
lined up at the gap threshold. 
In this case the inverse  square  root singularity  in  the density  of 
 states at  the  gap  threshold is  important and  the  adiabatic 
 approximation Eq.(\ref{gggg}) breaks  down. 

For shorter pulses the peak becomes
broader and is centered at larger $\omega$'s. If the integration range is
small, the result is quite sensitive to the location of the peak 
(see full line in Fig. (\ref{fig1})): most  remarkably,   
a  minimum appears  in the curve
when the pulse is rather long ($\omega_c/\omega_D<5$).
By contrast, the gap correction is rather 
flat w.r.to changes of $\omega_c$ when $\omega_D / \Delta$ is larger 
(broken and dotted line in Fig.\ref{fig1}).

\subsection{Correction  to the Josephson  current}
In this
subsection we derive the correction to the Josephson current arising
from the two terms of the anomalous propagator $f(R,t,t')$ given by
Eq. (\ref{anomalo}).
Eq.s(\ref{gggg},\ref{anomalo}) show that the  non equilibrium Keldysh Green
functions of the irradiated superconductor can be separated into two
terms. The first one is what we call the 'adiabatic' contribution,
while the second one is strongly dependent on the non equilibrium qp's
distribution function and is first order in  $q(t)$.
\\Within the linear response theory in the tunneling matrix element 
$|T_o|^2$, the pair current at zero voltage is:
\bea 
  J (\vec R  =0,t)=2e |T_o|^2 \int_{-\infty}^{\infty}\;dt'
 e^{2ieV(t-t')}
 \nonumber\\
 \left ({f^>}_1^\dagger (0,t, t')\: f^{A}_2(0,t',t) +
{f^R}^\dagger_2 (0,t,t') \: f^<_1 (0 ,t',t)
 \right )\;,
\label{jcur}
\enea
where $|T_o|^2$ is assumed to be independent of energy, for
simplicity. The current of Eq. (\ref{jcur})is evaluated at the junction
site, defined by $ \vec R = 0 $ and the irradiated
superconducting layer $S1$ is labeled by 1 here, while the
superconductor unexposed to radiation $S2$ is labeled by 2.  The
perturbed Josephson current has an adiabatic term  $J^{ad}(t)$ 
obtained by
inserting the first term of Eq. (\ref{anomalo}) into Eq. (\ref{jcur}),
plus a correction $\delta J(t)$
arising from the second term of Eq. (\ref{anomalo}). Using the
definitions ${f}^K = {f}^> + {f}^< $ and $ {f}^R - {f}^A = {f}^> -
{f}^< $ and expanding to lowest order in $q(t)$ the adiabatic critical
current is:
\bea
J^{ad}=J^{ad}_c(t) \sin(\varphi),\;\;\;\;\;\;
J^{ad}_c(t)=
\frac{\pi \h\Delta  }{2 e R_N}\left(1 + \frac{\delta \DD(t)}{2\DD} 
\right)\:,
\label{jos}
\enea
where $R_N$ is the normal resistence, 
$\Delta$ is the unperturbed gap parameter of both contacts
(assuming $\Delta_1=\Delta e^{i \varphi}$ and $\Delta_2=\Delta$ in the
absence of laser perturbation), and $\delta\Delta(t)/\Delta$ is given by
Eq. (\ref{delgap}).  Denoting by $\delta f^K$ the second term of
Eq. (\ref{anomalo}) the correction $ \delta J(t)$ is:
\beq
 \delta J(t) = 
e |T_o|^2 \int_{-\infty}^{\infty}\;d\ww  \:
\left ( (\delta  {f}^K(\omega+i e V)  )^\dagger _1
  {f}^A_2(\omega- i e V)  + ({ {f}}^R(\omega+ i e V))^\dagger _2  
 ( \delta  {f}^K  (\omega- i e V))_1 \right )\:.\label{newjos}
\eneq
At zero temperature and  $V = 0 $, this gives: 
\bea
\delta J(t)=\frac{\pi \h }{e R_N}  \int_{-\infty}^{\infty}\;d\ww \:
\frac{|\Delta|^2 q(t)}{\ww}  \rho\left [ 2(\ww ^2 -|\DD |^2 )^\met /(\sqrt {2}\ww
 _c ) \right ]\nonumber
\\ \left(
\frac{e^{-i\varphi}}{
\sqrt{(\omega+i0^+)^2-|\Delta |^2}}+\frac{e^{i \varphi}}{\sqrt{
(\omega-i0^+)^2-|\Delta |^2}}\right)  \label{cosphi} \:  .
\enea
which is zero for parity. This conclusion holds because we assume that
no charge imbalance occurs. If $V \neq 0 $ the contributions to the
integral evaluated in the complex plane are finite. \\$\delta J(t)$ is
a $cos \varphi$-like correction. In the unperturbed Josephson effect a
'$cos \varphi $' term only arises when  $V>2 \DD$ \cite{barone}. 
By contrast, our
calculation shows that a $cos \varphi$ term can arise in the Josephson
current with a small nonzero voltage in the presence of  an ultrafast
laser pulse.

\section{Classical dynamics of the irradiated junction}
In this section we integrate the classical equation of motion of the
irradiated junction numerically.  Here we discuss the possibility that
the laser pulse induces switches of the junction from the zero voltage
state, to the resistive state.  The characteristics of the Josephson
junction for a finite voltage, is obtained within the RCSJ
(resistively and capacitively shunted junction) model \cite{barone}.
The phase of the superconductor S2 is taken as the reference phase.

In the absence of the pulse, the junction is biased by a current
constant in time $J_{b}$. As discussed in the previous Section, the
pulse activates the superconductor S1 by varying its gap dynamically
in time.

Consequently, a voltage $V$
arises at the junction, related to the dynamics of the phase
difference $\varphi (t) $. The latter solves the differential equation:
\bea
\ddot  \varphi + Q_0^{-1} \dot \varphi  +  \frac { J_c^{ad}(t)}{J_c^o} 
 \sin \varphi(t)= \gamma
 \label{mmm}\; ,
\enea  
where $\gamma={J_{b}/J_c^o}$ and $J_c^o={(\pi\hbar \Delta)/(2eR_N)}$
is Josephson critical current of the unperturbed junction.  The
time-dependent driving term is deduced from Eq. (\ref{jos}).  We
assume $\omega_{pJ0} > \tau_E^{-1}$ ($\tau_E^{-1}(Nb)\sim 7 GHz$), where
$\omega_{pJ0}$ is the plasma frequency at zero bias. This condition is
satisfied for high quality $Nb$ junctions, where $\omega_{pJ0}$ is in
the range $40 GHz \div 120 GHz$ \cite{wallraff}, but it holds also if
we take into account the dependence of the plasma frequency on $\gamma
$: $\omega_{pJ}=\omega_{pJ0} (1 -
\gamma^2)^{(1/4)}$.  At $\gamma = 0.98$ the term $(1
- \gamma^2)^{(1/4)} = 0.44$: this still gives a large plasma frequency
for the given range. In any case the plasma frequency changes
marginally when the energy is degradated into heat if $q$ is
small. Under this conditions the relaxation process occurs long after
the switch to resistive state.

In Eq. (\ref{mmm}), $Q_0=\omega_{pJ0} R(\varphi) C$ is the quality
factor, where $R(\varphi)$ is the junction intrinsic resistance, which
is in general a non-linear function of the phase. 
The dissipative $Q^{-1}_0\dot
\varphi$ term includes thermal incoherent pair breaking effects at
equilibrium.  In the simulation we use both a constant junction
resistance $R$ and a patchwork model given by
\cite{Likharev}:
\begin{equation}
Q^{-1}(\dot\varphi)=Q_0^{-1}\frac{\omega_{pJ}}{\Delta}\frac{(\frac{\dot
\varphi \omega_{pJ}}{\Delta})^{N}}{1+(\frac{\dot \varphi \omega_{pJ}
}{\Delta})^{N}}
\label{NLD}
\end{equation}
with $N=16$ and $Q_0^{-1}=0.636$, which corresponds to a normal
resistance above the gap $R_N=(\omega_{pJ} \varphi_0)/(c J^0_c)$. In
general we ignore the direct dependence of $R(\varphi)$ on the
phase. By the way, $Q$ should also depend on the energy which is
released by each single laser pulse due to the incoherent pair
breaking process. However, under the hypothesis that this energy is
very low we assume that the quality factor, due to the optically
induced normal resistance of the sample, is constant within the
considered energy range.\\ Actually, in the presence of the pulse,
also the current contribution of Eq. (\ref{newjos}) should be plugged
into the l.h.s. of Eq. (\ref{mmm}).  This current term depends on the
voltage $V=\dot \varphi$. However, in view of the fact that in this
work we are only concerned with the switching of the junction out of
the zero voltage state, we do not derive the full dynamics of the
phase self-consistently.

\begin{figure}[ht]
   \vspace{0.5cm} \hspace{-0.5cm} \includegraphics[width=\columnwidth
   ]{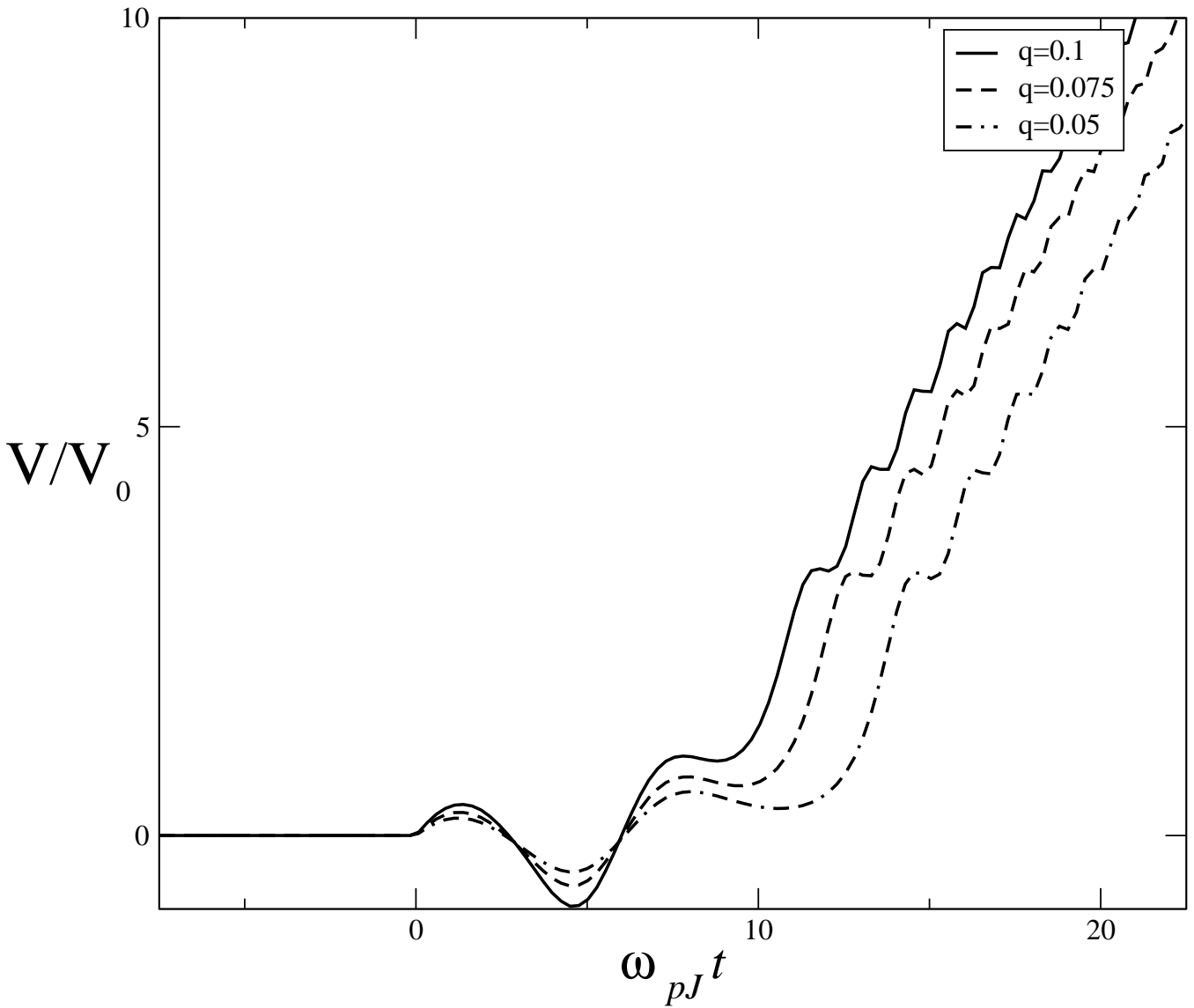} \caption{Voltage behavior in time for
   different energy released on the sample and the quality factor $Q =
   100$. The voltage is normalized to $V_0=\omega_{pJ} \varphi_0/c$.}
\label{voltaggio}
\end{figure}
\begin{figure}[ht]
   \vspace{0.5cm} \hspace{-0.5cm} \includegraphics[width=\columnwidth
   ]{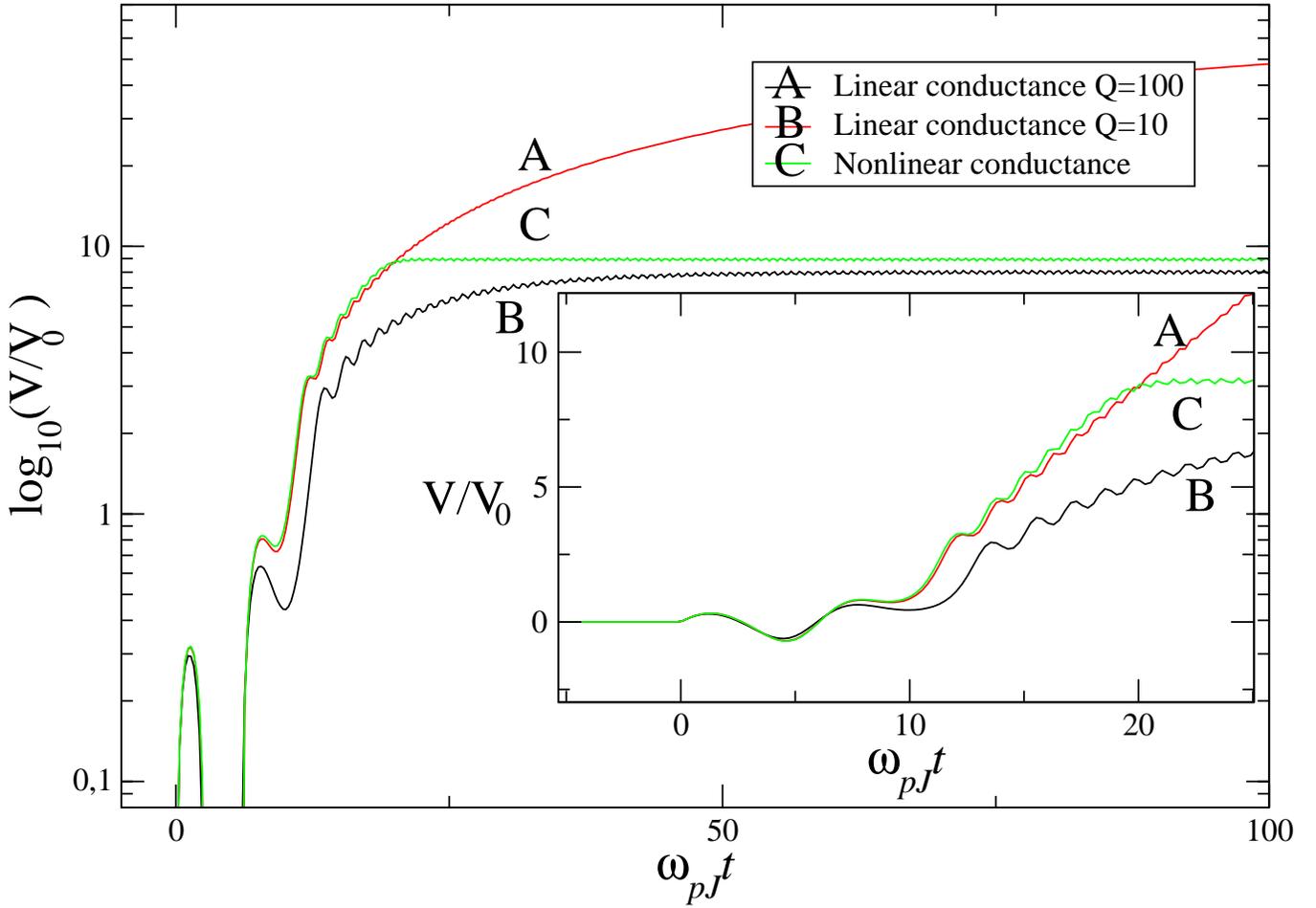} \caption{(Color online) Voltage behavior in time with different
   quality factors $Q$ in the linear conductance model (A,B) 
   and the non-linear conductance model as given by
   Eq. (\ref{NLD}) (C). The voltage is normalized to $V_0=\omega_{pJ}
   \varphi_0/c$.}
\label{voltaggio1}
\end{figure}
\begin{figure}[ht]
  \vspace{0.5cm} \hspace{-0.5cm} \includegraphics[width=\columnwidth
  ]{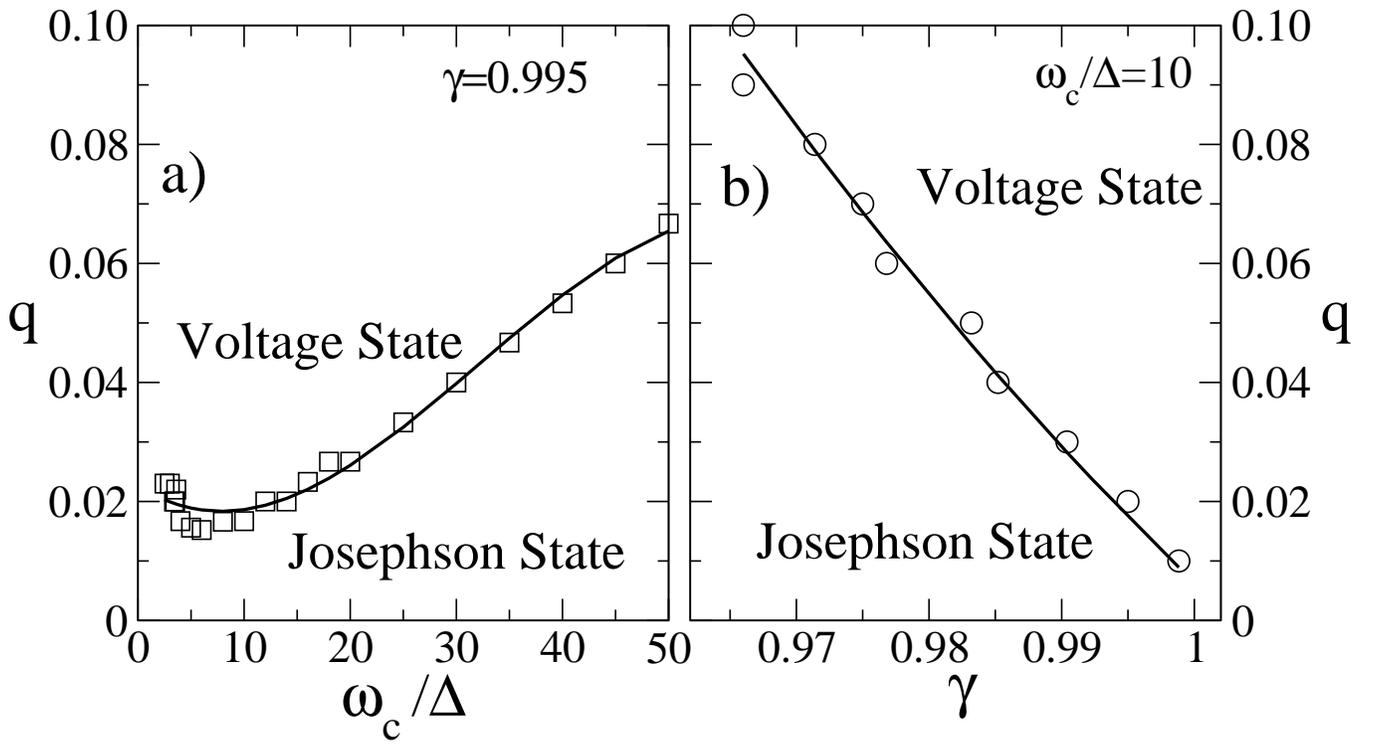} \caption{Switching front in the parameter space
  a)$(\omega_c/\Delta,q)$ at fixed current bias $\gamma  =  J_b/J^o_c $ and b)
  $(\gamma,q)$ at fixed $\omega_c/\Delta$, for $Q = 100$ and
  $\omega_D/\Delta=10$.  The full curves have been added as a guide
  for the eye: they mark the border between the zero voltage
  (Josephson) state and the voltage state. $q$ is the coupling
  strength due to the laser pulse.}
  \label{fig5}
\end{figure}

In Fig. (\ref{voltaggio}) we show the voltage just after the pulse for
different values of the released energy.  The time evolution of the
voltage is sketched for some successfully induced switches. The
junction starts in the zero voltage state. At $\omega_{pJ}t=0^+$ it is
irradiated by the laser pulse.  There are few oscillations at
frequency $\omega_{pJ}$ before the switching occurs, followed by an
overall increase of the oscillating voltage.  The larger is $q$ the
faster is the switch.  If no switch is induced the junction remains in
the zero voltage state: the phase and the voltage are weakly perturbed
by the radiation and show decaying plasma oscillations around their
equilibrium values.

In Fig.(\ref{voltaggio1}) we show the approach to the gap voltage for
different $Q$ values and two different conductance models. Except
for the asymptotic trend, the curves for $Q=10$ (B) and $Q=100$(A) show a
similar behavior. The non-linear conductance gives rise to a more
pronounced shoulder in the curve after the first increase of the
voltage. The first phase oscillation at frequency $\omega_{pJ}$ are
largely independent of the dissipation model used.

The switching of the junction out of the zero voltage state depends on
the bias current $J_{b}$, on the released energy per pulse $q$, and on
the pulse duration. In Fig.(\ref{fig5}) we sketch the switching front
in the parameter space a)$(\omega_c/\Delta,q)$ at fixed $\gamma$ and
b) $(\gamma,q)$ at fixed $\omega_c/\Delta$, for $Q = 100$ and
$\omega_D /\Delta=10$.  For each point $(q,\omega_c/\Delta ,\gamma)$
we calculate $J_c^{ad}(t)$ from Eq.s(\ref{jos}, \ref{delgap}). Next we
plug the result into the equation of motion Eq. (\ref{mmm}). Numerical
simulation of the dynamics shows whether the junction is stable in the
zero voltage state, or it switches to a running state. The points of
the curve mark the frontier between the two behaviors. The full curve
is just a guide for the eye.  The non-monotonicity of
$\delta\Delta(t)/\Delta$ with the pulse duration, appearing in
Fig.(\ref{fig1}) forces a similar behavior in
Fig.({\ref{fig5}}$a$). This means that the pulse duration can be
appropriately chosen, in order to optimize the junction switching with
the laser field. Indeed, if $\omega_c/\Delta$ is of the order of
$5\div 15$ a very small released energy is required for the switching
of the junction, because the order parameter is much depressed by the
laser pulse in that range. Out of this range the shorter is the pulse
the larger is the energy required. By contrast a slightly larger
released energy is also required for longer pulses
($\omega_c/\Delta<5$). This is because longer pulses imply a more
extended change in the qp distribution up to higher energies. As a
consequence the maximum of the function $\rho$ of Eq.
\ref{delgap} contributes less to  the correction of the gap parameter.
Nevertheless caution should be used in considering our results for
longer pulses because of the neglected relaxation effects.

\section{Conclusion}
The effect of an ultrafast laser pulse on the superconducting
coherence at a Josephson Junction allows studying an unexplored regime
in non equilibrium superconductivity.

Non-equilibrium in superconductivity is usually addressed in the
context of one of the possible applications of Josephson junctions,
that is radiation$/$particle detectors. Highly energetic radiation
produces pair breaking and quasiparticles which, in turn, excite a
large number of them, in a cascade process.  Usually the setup is
optimized such that the qp's can be collected and contribute to the
current across the junction with a sharp signal. Losses are due to
degradation of the released energy into heat during the relaxation
process.  To achieve optimum performance, the Josephson current is
usually suppressed by applying a magnetic field. This picture has been
discussed quantitatively by examining the quasiclassical kinetic
equation for the non equilibrium qp distribution function
\cite{ovchinnikov}.

In this work we have concentrated on a quite different time scale: the
one fixed by the duration of an ultrafast laser pulse. While the
relaxation process mentioned above takes place on a time scale of $0.1
\div 100 \; ns$, we have considered a laser perturbation lasting at
most hundreds of femtoseconds. This type of tool can be quite valuable
for future applications, because fast pulses in flux and gate voltages
are extremely important when processing information in superconducting
quantum computing devices (qubits) \cite{makhlin}. Indeed, finite rise
and fall times of pulses may result in a significant error in
dynamical computation schemes \cite{choi}. The carrier frequency of
the laser is $\sim 100 THz$ and the optical radiation is expected to
produce many $e-h$ pair excitations as it would occur in a normal
metal.  In our case, qp's do not have enough time to relax down to the
typical phonon frequencies ($\sim \omega_D$) and to heat the sample
before the stimulated switching occurs. We do not wish to collect qp's either,
what requires a suitable geometry of the junction.

Instead, we have addressed the question how the critical current for
Josephson conduction $J_c$ can be coherently affected by a laser
induced small perturbation with an unrelaxed non-equilibrium
distribution of qp's, that is before the dissipative response sets
in.

Using the quasiclassical approach to non-equilibrium Keldysh Green's
functions, we have shown that, if the temperature is very low, the
order parameter of the irradiated superconductor can respond
adiabatically to a weak perturbing signal. A non equilibrium
distribution of qp's is generated and consequently $J_c$ is
temporarily reduced (see Eq. (\ref{jos}) and Fig(\ref{fig1})).  This
reduction can drive switches of the junction out of the zero voltage
state.  In our approach the retardation effects which arise from the
frequency dependence of the $e-ph$ coupling $\alpha^2 F(\omega)$ and
from the actual features of $e-ph$ relaxation processes have been
neglected.  They came into play on a time scale longer then the
duration of the laser pulse, $\omega _c ^{-1} \sim \tau _c $.  Indeed
in the equation of motion for the Keldysh Green's function reported in
appendix B and C the role of the frequency dependent self-energy terms
has not been discussed.

The parameter which is
related to the energy released by the radiation and describes the
strength of the perturbation is $q$.  In our case, the switches can be
induced by pulses with $q \sim 0.05$, with relatively low values of
$\omega_c/\Delta\sim 10$, by polarizing the junction very close to the
critical current.

In experiments on laser induced non-equilibrium effects in
superconductors \cite{testardi} or Josephson junctions \cite{parker},
the released energy is of few $\mu J$, which corresponds to values of
$q$ between $0.25$ and $0.67$ for the given laser spot dimension.  
In our case for $q=0.05$ the switching occurs at 98\% of the critical
current. Therefore,  a coherent effect of the laser on the 
superconducting condensate is sufficiently  large to be observed in sensitive 
experiments monitoring the  escape rate \cite{thermal,quantum}. 
These experiments can appreciate very  small variations of the critical 
current, if temperature is kept low and   the released energy is 
sufficiently small, so that sizeable heating effects do not occur.

We have also found a `$\cos \varphi $' contribution to the Josephson
conduction due to the presence of the excited qp's (see
Eq. (\ref{cosphi})).  This term, which will be examined in detail
elsewhere, vanishes at zero $V,T$, provided excitations do not
generate charge imbalance. A similar term, proportional to the voltage
$V$, can be derived also in the BCS theory of Josephson conduction
\cite{barone} but it is identically zero as long as $V \le 2 \Delta /e
$, because of the absence of qp's at zero temperature. This is not the
case here, due to the presence of a non-equilibrium qp distribution.

Our derivation of Eq.s(\ref{delgap},\ref{jos},\ref{cosphi}) assumes
that no charge imbalance is created by the perturbation. This is
because the radiation excites $e-h$ pairs and the pulse duration is
short enough, so that pair breaking is very limited.  The absence of
charge imbalance is a crucial approximation in our solution
scheme. This assumption allows us to keep the unperturbed functional
form of the quasiclassical Green's functions and to insert a time
dependent gap parameter $\Delta (t)$ in their expressions. $\Delta
(t)$ follows the perturbation adiabatically and is determined by the
non equilibrium $e-h$ pair distribution produced by the pulse. Charge
imbalance corrections should be reconsidered, but they are usually
expected to have a minor effect on the dynamics of the junction. 

To complete the picture, we have simulated the classical dynamics of
the junction switching to the resistive state. This picture is only
valid over few periods $2 \pi /\omega_{pJ}$ on time duration less than
the electron-phonon relaxation time.

Precursor oscillations can be seen  in Fig.  (\ref{voltaggio})
at the Josephson plasma frequency $\omega _{pJ}$. 
Most remarkably the duration of the pulse can be optimized 
in order to induce controlled coherent switching 
at the minimum possible released energy ${\cal{E}}$. 

\begin{acknowledgments}
We are indebted to B.Altshuler, A.Barone,  D.Bercioux, L.N.Bulaevskii,
F.Hekking,  Yu.N.Ovchinnikov,  G.Pepe and  G.Schoen for useful discussions
 at various 
stages in the preparation of this paper.
\end{acknowledgments}
\appendix
\section{Quasiclassical time dependent  Green's function approach}
The quasiclassical Green's function solve the Eilenberger form of the
Gorkov equations in commutator form \cite{belzig}: 
\beq 
\left [ \left
(\breve{G}_0^{-1} -\breve \Sigma -\breve \Delta \right ),
\GG \right ]_ \otimes =0 \;.
\label{gorkov}
 \eneq 
The matrix Green's function $\GG$ in the Keldysh space is: 
\beq
 \GG= \left( \begin{array}{cc}
\hat G^R & \hat G^K\\
\hat 0 &\hat G^A
\end{array}\right) \:\: ,
\eneq
where, in  turn,    $\hat G^R,\hat G^A,\hat G^K$  are the retarded,
advanced and Keldysh   Green functions in the Nambu space:
\beq
\hat G^{(A,R,K)} =
\left( \begin{array}{cc}
 g^{(A,R,K)} &  f^{(A,R,K)}\\
 -f^{(A,R,K)\dagger}  & - g^{(A,R,K)\dagger}
\end{array}\right)\:\;.
\eneq
Here $f$ is the anomalous propagator and   its Keldysh component
defines the gap:
\beq
\Delta= \frac{\nu (0)\lambda}{4} \int_{-\infty}^{+\infty} d \omega\; <f^K>_{\vec v_F} 
\:.  \label{eqgap}
\eneq
The average $<>_{\vec v _ F}$ denotes angular averaging over the Fermi
surface.
The gap matrix $\breve \DD$ is defined as:
\beq 
\breve \Delta \equiv  \left(
\begin{array}{cc}
\hat  \Delta  & 0\\ 0 &\hat \Delta 
 \end{array}\right) \: ,
\:\:\:\:
\hat \Delta= \left(
\begin{array}{cc}
0&  -\Delta  \\  \Delta^* &0
 \end{array}\right) \: . \label{eqgap1}
\eneq
The self-energy $\breve \Sigma $ includes elastic and inelastic
scattering with impurities and gives rise to relaxation processes.
The commutator is evaluated w.r. to the $\otimes$ operation, which
implies integration over the intermediate variables according to:
\beq
\GG _0^{-1}\otimes \GG _0\equiv \int d2\: \GG _0^{-1}(1,2)
 \GG (2,1') \:  ,
\eneq
where   $1 \equiv (\vec r_1 ,t_1)$.
The differential operator $\GG _0^{-1}(1,2)$ is
\beq
\GG _0^{-1}(1,2)=\left [ i\breve \sigma _3 
\partial _{t_1} -  
\frac{1}{2m}  \left (\vec \nabla_{\vec{r}_1}  -i \frac{e}{c} \breve {\sigma }_3
 \vec A(1) 
 \right ) ^2 +(e\phi(1) -\mu) \breve I  \right ]\delta(1-2)\;.
\eneq
Here
\beq 
\breve \sigma_i \equiv  \left(
\begin{array}{cc}
\hat  \sigma_i  & 0\\ 0 &\hat \sigma_i
 \end{array}\right) ,\;\;\;\;\breve I \equiv  \left(
\begin{array}{cc}
\hat  I  & 0\\ 0 &\hat I
 \end{array}\right) \;\;\;\;,
\eneq
where $\hat{\sigma}_i (i=1,2,3)$ are the usual Pauli matrices and $\hat I$ the 
$2\times2$ unit matrix .\\
The vector potential $ \vec A(1) $
 describes the laser radiation field of frequency
$\Omega$: 
\beq 
\vec{ A}(1)=\sum_{\pm} \sum  _{\vec{p}}\vec{a}_\pm
(p,t)e^{\mp i (\vec{p}\cdot \vec{r}_1 - \Omega t)} \; , 
\eneq
where $\vec{a}_{\pm}$ can be  slowly varying 'envelope' functions of 
 space,   on the
light spot size $ R_o $ and on   time, on  the  scale $\Omega^{-1}$.  These
are the reference space and time scales in the following.
In the frame of the quasiclassical approximation, the original Green's
functions $ G(1,2) \equiv G(\vec{r},t, \vec{r}',t') $ are
assumed to be slowly varying function of the coordinate $\vec R
=(\vec{r}+\vec{r}')/2 $ while they oscillate fast as functions of
$\vec{r}-\vec{r}'$ on the scale of the Fermi wavelength $\lambda_F$.
It is customary to rewrite also the time dependence in terms of the
new variables $ \overline t = ( t +t')/2 $ and $ t-t' $ and to Fourier
transform $w.r.to$ $\vec{r}-\vec{r}'$ and $ t-t' $, thus obtaining $
G( \vec{p},\omega ,\vec{R},\overline t ) $.\\
The motion equation for the Keldysh component of
eq.(\ref{gorkov})  reads:
\beq
\left[ \hat G_0^{-1} -\Re e (\hat  \Sigma) 
-\hat \Delta,\hat G^K \right ]_{\otimes}
 = \left[ \hat \Sigma ^K,\Re e ( \hat G)
\right]_{\otimes} +\frac{i}{2}\left
\{\hat \Sigma^K ,\hat A     \right\}_{\otimes}
-\frac{i}{2}\left\{\hat \Gamma , \hat G^K\right\}_{\otimes} \:  ,
\label{kcomp}
\eneq
where we have defined a quantity proportional to the density of states
$\hat A= i(\hat G^R-\hat G^A)$ (not to be confused with the vector
potential), and written down the imaginary and the real part of the
self-energy, $\hat\Gamma=i(\hat\Sigma^R-\hat\Sigma^A)$ and $\Re e
\hat\Sigma=\met (\hat \Sigma^R+\hat \Sigma^A)$, respectively, as well
as the  real part of the retarded/advanced Green's function $\Re e
\hat G =\met (\hat G^R+\hat G^A)$  ($\{ ,\} $ denotes the anticommutator ).
The next step is the gradient expansion of the $\otimes$ product (we
drop the overline on $t$ in the following):
\beq 
\hat C \otimes \hat B = \exp {1/2 (\partial^C_t \partial^B_\omega -
\partial^C_\omega  \partial^B_t) } 
\exp {1/2 (\partial_p^C \partial_R^B-\partial^C_R \partial_p^B )}
 \hat C \hat B \:,
\eneq
(here $\partial_p^C$ stands for the gradient w.r.to the impulse
operating on $\hat C$ ), followed by the averaging of the result for
$| \vec{p}|$ close to $p_F$, that is over the energies $p^2/2m-\mu
\equiv\xi $ while the direction of $\vec{p}$, $\hat p $, is untouched:
\beq
\hat{g}(\hat{p},\omega ,\vec{R} , t)=\frac{i}{\pi}\int
d\;\xi \:\:  \hat{G}(\vec{p}, \omega ,\vec R , t) \:\:  .
\eneq
This is justified, because $\lambda_F$ is much shorter both of the
superconducting correlation length and of the spatial range of the
laser spot ($e-h$ symmetry is assumed).  Eventually $\hat g$ depends on
$\hat p,\omega,R, t$.  The average of eq.(\ref{kcomp}) over all
directions in the Fermi surface, can be done if no external bias is
applied and anisotropies of the diffusion and relaxation process are
not expected.  In the presence of radiation with optical frequency it
is customary to average out the fast oscillating components with
frequency $\Omega $
\cite{landau}.  Following eq.(\ref{potvet}), we expand the Green's
functions similarly:
\beq
 \hat{\underline g}(\omega,R,t) = {\hat g} (\omega,R,t) + \sum_{\pm}
\hat{\chi} ^{\pm}
 (\omega,R,t) e^{\pm i \Omega t}\; . 
\eneq 
Here $\underline {\hat g }(\omega,R,t)$ is assumed to be the slowly
varying part on the scale of the pulse duration $\tau _c$, while $\hat
{\chi}^\pm $ are fast varying ones.  All these functions are slowly
varying functions of space as well, on the light spot size scale.  A
decomposition of Eq. (\ref{kcomp}) into harmonics arises.  We are
interested in the zero order harmonic equation, which shows a slow
dynamics that can be followed coherently by the irradiated
superconductor. Leaving the self-energy terms in Eq. (\ref{kcomp}) for
the moment aside and dropping the superscript $^{(K)}$, we obtain
\beq
[\tau_3 a_+,\hat {\chi}^-]+[\tau_3 a_-,\hat {\chi}^+]
+[(\omega \hat \tau_3 - \hat \Delta),  \ug]
-\met\{\hat\tau_3,\partial_t  \ug\} +\met  \{\partial_t
\hat \Delta,\partial_\omega  \ug\}-\frac{e^2}{2mc^2}
\partial_t A^2\partial_\omega   \ug = ... \: ,
 \label{gorfin}
\eneq
where the ellipsis refers to the missing self energy terms. The first
order harmonic equations are:
\beq
  \pm i \; \{\tau_3, \hat{\chi}^{\pm}\} +\frac{ e }{2mc \Omega}[a_\pm,\ug]=0\;,
\eneq
they show that the first two  terms in  Eq. (\ref {gorfin}) are ${\cal{O}} (
\Omega ^{-1}) $  smaller than the others and can be neglected to  lowest 
order. Hence the effective equations for the Green functions are 
\beq
[(\omega \hat \tau_3 - \hat \Delta),  \ug]
-\met\{\hat\tau_3,\partial_t  \ug\} +\met  \{\partial_t
\hat \Delta,\partial_\omega  \ug\}-\frac{e^2}{2mc^2}
\partial_t A^2\partial_\omega   \ug = ... \: ,
\label{gorfin2}
\eneq
The last three terms in Eq. (\ref {gorfin2}) include the time
dependent non equilibrium dynamics that is absent in the case of a time
independent approach. 
Retarded, advanced and Keldysh  Green's function, they all
 satisfy analogous equations.
\section{Kinetic equation  for $n(\omega,r,t)$} 
One can linearize Eq. (\ref{gorfin2}) for Keldysh Green's function by
posing $\hat g^K=\hat g^R \hat h - \hat h \hat g^A$. This yields the
kinetic equation for the distribution function $n(\omega, R,t)$
\cite{scalapino}. We
neglect any variation in space and concentrate on the $t$ dependence
here.  The distribution matrix $\hat h$ is defined starting from the
$n_L$ and $n_T$ functions according to:
\beq 
\hat h= n_L \hat 1 + n_T \hat \sigma_3\:,
\eneq  
or 
\beq \hat h =
\left( \begin{array}{cc}
n_L (E)+n_T(E)& 0\\
0&n_L (E) - n_T(E) \end{array}\right) \equiv \left( \begin{array}{cc}
1-2n(E)& 0\\
0&2n(-E)- 1\end{array}\right)\:   .
\label{funh}
\eneq
The  second equality  defines  the  relation with  the  qp  distribution 
 function  $n (E)$. We  always  assume  $e-h$ symmetry, so  that  
$n(E)+n (-E ) =1 $ and $n_T = 0 $. 
In the equilibrium case  one has  $ n_o(E)=\frac{1}{e^{E/T}+1} $,  so  that:
\beq
n_{L(T)}^o= \met\left[tanh(\frac{E}{2T})+(-)tanh(\frac{E}{2T})\right]\:.
\eneq
  Substituting Eq. (\ref{ggk})
in Eq. (\ref{gorfin2}), we get the kinetic  equation for the distribution
matrix $\hat h $.In particular, in case there is no charge imbalance,
 the  equation for the longitudinal component $n_L$ is
\cite{ovchinnikov}:
\beq
 \partial_ t n_L Tr((\hat g ^R \hat \tau_3- \hat \tau_3 \hat g^A))+
 \partial _\ww n_L Tr(\partial _t \hat \DD (\hat g ^R - \hat g^A))
-\frac{e^2}{2 m c^2}\partial_t A^2 Tr
 \left(\partial_\omega n_L\left(\hat g^R-\hat
 g^A\right)\right)
 =
 -4 I[n_L(\ww)] \:, \label{ovA2}
\eneq
where $I[n_L(\ww)]$ is a collision integral which regulates the
relaxation of the qp distribution toward equilibrium. Eq. (\ref{ovA2}) 
is
fully discussed in ref.\cite{ovchinnikov}.

\section{Equation of motion of  $ G^R $}
We now write down the equations Eq. (\ref{gorfin2}) explicitly for the
retarded Green's functions. We label the matrix components by $(i,j)\:
(i,j=1,2 )$ and drop the superscript $^R $ everywhere.  The matrix
elements of Eq. (\ref{gorfin})in the Nambu space become :
\bea
&{\bf (1,1)\to }& \partial_t g -  \DD f^\dagger + \DD^* f
   +  \partial_t \Delta \partial_\omega f^\dagger+\partial_t \DD^*
\partial _\omega f
  +  \frac{e^2}{2 m c^2} \partial_t A^2 \partial_ \omega g= ...\nonumber\\
&{\bf (2,2) \to }&  \partial_t g^\dagger
 + \DD f^\dagger- \DD^* f 
 + \partial_t \Delta \partial_\omega f^\dagger+ \partial_t \DD^*
\partial _\omega f 
 -\frac{e^2}{2 m c^2} \partial_t A^2 \partial_ \omega g^\dagger =
 ...\nonumber\\ 
&{\bf (1,2) \to }& 2 \omega f - \DD g^\dagger- g \DD
 + \partial_t \DD \partial_\omega g^\dagger - \partial_\omega
 g
\partial_t \DD  + \frac{e^2}{2 m
c^2}\partial _t  A^2
\partial _\omega f= ...\nonumber\\
 &{\bf (2,1) \to }&   +2 \omega f^\dagger - \DD^* g - g^\dagger \DD^*
  +  \partial_t \DD^* \partial_\omega g - \partial_\omega 
g^\dagger
\partial_t \DD^*
 - \frac{e^2}{2 m c^2} \partial _t A^2 \partial _\omega f^\dagger	= ...
\nonumber\:\:\:\:\:\:.
\enea
The equilibrium result suggests that
\beq
 f=\frac{\Delta(t)}{\omega} g \label{ff}
\eneq  
solves Eq.$(1,2)$ except for terms $\propto\partial_t A^2$ which
describe the relaxation at later times. Let us assume that this
relation holds also in the non-equilibrium case.  Then the formal
solution, follows adiabatically the $t-$dependence of the gap parameter $\Delta$ by keeping an equilibrium-like shape:
\beq
g_{ad}=\frac{\omega}{\sqrt{\omega^2-|\Delta|^2(t)}},\;\;\;\;\;\;
f_{ad}=\frac{\Delta(t)}{\sqrt{\omega^2-|\Delta|^2(t)}}
 \label{fg}
\eneq 
This approximate solution is quite appealing, because it satisfies 
the equilibrium
condition for $\Delta$ at $t\to \infty $. However it neglects
diffusion in space-time which will be mainly important at later times
w.r. to the pulse duration.

%
%


\begin{thebibliography}{40}
\bibitem{barone} A. Barone and G. Patern\'o, 
{\it Physics and applications of Josephson effect}, Wiley New York,
1982.
\bibitem{testardi} L. R. Testardi, Phys. Rev. B {\bf 4}, 2189, 1971.
\bibitem{parker} W. H. Parker and W. D. Williams, Phys. Rev. Lett. {\bf 29}, 924, 1972.
\bibitem{owen} C. S. Owen and D. J. Scalapino, Phys. Rev. Lett. {\bf 28}, 1559, 1972.
\bibitem{ivlev} R. A. Vardanyan and B. I. Ivlev, Sov. Phys. JEPT {\bf 38},
1156, 1974.
\bibitem{ser}A.V. Sergeev, M.Yu. Reizer,
J. Mod. Phys. B \textbf{6} 635 (1996), M.Yu. Reizer, Phys. Rev.
B \textbf{39} 1602 (1989), A.V. Sergeev, M.Yu. Reizer, {Zh. Eks.
Teor. Fiz.} \textbf{90} 1096 (1986) [{Sov. Phys. JETP} \textbf{63}
616 (1986)]
\bibitem{lindg} M. Lindgren, M. Currie, C. Williams, T.Y. Hsiang,
P.M. Fauchet, R. Sobolewski, S.H. Moffat, R.A. Hughes, J.S.
Preston, and F.A. Hegmann {Appl. Phys. Lett.} \textbf{74} 853
(1999)
\bibitem{semenov}A.D. Semenov, R.S. Nebosis, Yu. Gousev, M.A.
Heusinger, K.F. Renk, {Phys. Rev. B} \textbf{52} 581 (1995).
\bibitem{kaplan}S.B.Kaplan, C.C.Chi, D.N.Langenberg, J.J.Chuang, S.Jafarey, 
D.J. Scalapino, Phys.Reb.B {\bf 14}, 4854, 1976.
\bibitem{thermal} M. H. Devoret, J. M. Martinis, D. Esteve and J. Clarke, 
Phys. Rev. Lett. {\bf 53}, 1260, 1984.
\bibitem{quantum} M. H. Devoret, J. M. Martinis and J. Clarke, Phys. Rev. Lett. {\bf 55}, 1908, 1985.
\bibitem{latching} A.Moopenn, E.R. Arambula, M.J. Lewis, H.W. Chan, 
IEEE Trans. Appl. Supercond. 3, 2698, 1993,
\bibitem{pulse} G. P. Pepe, G. Peluso, M.Valentino, 
A. Barone, L. Parlato, E. Esposito, C. Granata, M. Russo, 
C. De Leo and G. Rotoli Appl. Phys. Lett. {\bf 79}, 2770, 2001.
\bibitem{barone2}Superconductive Particle Detectors, 
edited by A.Barone (World, Singapore, 1988); A. Barone, Nucl.Phys. B {\bf 44} 645 (1995). 
\bibitem{ovchinnikov}Yu.N.Ovchinnikov and V.Z.Kresin 
Eur.Phys.J.B {\bf 32}, 297, (2003), Yu.N.Ovchinnikov and V.Z.Kresin, Phys.Rev. B. {\bf 58},12416,(1998).
\bibitem{rammer} A.I.Larkin and Yu.N.Ovchinnikov Zh.Eksp.Teor.Fiz {\bf 55},2262 (1968) [Sov.Phys. JETP {\bf 28}, 1200, 1969];
J.Rammer, H.Smith, {Rev. Mod. Phys.},
\textbf{58}, 323,(1986).
\bibitem{belzig}W.Belzig, F.K.Wilhelm, C.Bruder, G.Schoen, E.D.Zaikin
Superlattices and Microstructures {\bf 25}, 1251 (1999).
\bibitem{amin}M.H.S.Amin, Phys.Rev.B {\bf 68 }, 054505, (2003).
\bibitem{noi}preliminary  report on some of this  material has  appeared on 
P.Lucignano, F.J.W.Hekking and A.Tagliacozzo,   in ``Quantum computing and Quantum  bits in  mesoscopic
  systems'', A.J.Leggett,  B.Ruggero, P.Sivestrini 
ed.s, Kluwer Academic, Plenum  Publishers N.Y. (2004).
\bibitem{grimvall} G.Grimvall 'The electron phonon interaction in metals', North Holland Publ. Co. Amsterdam 1981
\bibitem{scalapino}J.J.Chang and D.J.Scalapino, Phys.Rev.B {\bf 15 },
 2651, (1977).
\bibitem{eliashberg} L.P.Gorkov and G.M.Eliashberg,Soviet Physics,JEPT{27},328
(1968)
\bibitem{gork} A.A.Abrikosov, L.P. Gor'kov, I.E.Dzyaloshinski, {\it Methods of quantum field theory in statistical physics};
\bibitem{wallraff}A. Wallraff, A. Franz, A. V. Ustinov, V. V. Kurin,
I. A. Shereshevsky and N. K. Vdovicheva, PhysicaB {\bf 284-288}, 575, 2000.
\bibitem{Likharev} K. K. Likharev, {\it Dynamics of Josephson junctions and circuits}, Gordon and Breach Science Publishers, 1986.
\bibitem{makhlin} Yu. Makhlin, G. Sch{\"o}n, and A. Shnirman,
{Rev. Mod. Phys.} \textbf{73}, 357 (2001)
\bibitem{choi} M.S. Choi, R. Fazio, J. Siewert, and C. Bruder,
{Europhys. Lett.} \textbf{53}, 251 (2001)
\bibitem{landau} L.D.Landau, E.M.Lifsits, Mechanics, Pergamon
Press (1960).
\end{thebibliography}
\end{document}